\documentclass[preprint,onecolumn,showpacs,amsmath,amssymb,floatfix,pre]{revtex4}
\usepackage{amsfonts}
\usepackage{amssymb}
\usepackage{amsmath}
\usepackage{epsfig}

\begin{document}
\title{Quasi-stationary simulation: the subcritical contact process}
\author{Marcelo Martins de Oliveira}
\email{mancebo@fisica.ufmg.br}
\author{Ronald Dickman}
\email{dickman@fisica.ufmg.br} \affiliation{Departamento de F\'\i
sica, ICEx, Universidade Federal de Minas Gerais, 30123-970, Belo
Horizonte, Minas Gerais, Brazil} \draft

\begin{abstract}
We apply the recently devised quasi-stationary simulation method to
study the lifetime and order parameter of the
contact process in the subcritical phase. This
phase is not accessible to other methods because virtually all
realizations of the process fall into the absorbing state before the
quasi-stationary regime is attained.  With relatively modest
simulations, the method yields an estimate of the
critical exponent $\nu_{||}$ with a precision of 0.5\%.
\end{abstract}

\maketitle

\section{Introduction}

Stochastic processes with an absorbing state arise frequently in
statistical physics \cite{vankampen,gardiner}, epidemiology
\cite{bartlett}, and related fields.
Phase transitions to an absorbing state in spatially extended
systems, exemplified by the contact process \cite{harris,liggett},
are currently of great interest
in connection with self-organized criticality \cite{socbjp},
the transition to turbulence \cite{bohr}, and
issues of universality in nonequilibrium
critical phenomena \cite{marro,hinrichsen,odor04,lubeck}.

Systems exhibiting a phase transition to absorbing an state possess
(for appropriate values of the control parameter), an active
(nonabsorbing) stationary state in the infinite-size limit. But if
the system size is finite, the process must eventually end up in the
absorbing state. The {\it quasi-stationary} (QS) distribution for
such a system provides a wealth of information about its behavior.
(Since the only true stationary state for a finite system is the
absorbing one, simulations of ``stationary" properties of models
with an absorbing state in fact study the quasi-stationary regime.)

Conventional Monte Carlo methods entail a somewhat complicated
procedure for determining QS properties: many independent
realizations are performed (using the same initial configuration),
and the mean $\phi(t)$ of some property (for example the order
parameter) is evaluated over the surviving realizations at time $t$.
At short times times $\phi(t)$ exhibits a transient as it relaxes
toward the QS regime; at long times it fluctuates wildly as the
surviving sample decays. In the supercritical phase (where an
infinite system survives forever) one is normally able to identify
an intermediate regime free of transients and with limited
fluctuations, which can be used to estimate the QS value of $\phi$.
(Even in this case the method requires careful scrutiny of the data
and is not always free of ambiguity \cite{heger}.)  Deep in the
subcritical phase, however, conventional simulations are
impractical, as nearly all realizations fall into the absorbing
state before the quasi-stationary regime is attained. (In other
words, one no longer has a separation of the time scales for
relaxation to the QS state and for its survival. The effective
potential governing the process no longer possesses a local minimum
away from the absorbing state.)

We recently devised a simulation method that yields quasi-stationary
properties directly \cite{qssim}.  We showed that the method
reproduces known scaling properties of the contact process (CP);
in \cite{qsfisica} it was used to study the static correlation
function of the model.  In studies of the critical point, the QS simulation
method requires an order of magnitude less computer time than
conventional simulations, to obtain results of comparable precision.
Here we use the method to study the lifetime and the order parameter
of the one-dimensional contact process in the subcritical phase.

In the following section we review the method and define the contact
process. Then in Sec. III we present our results for the lifetime of
the CP on a ring, in the subcritical phase. We summarize our
findings in Sec. IV.

\section{Background}

We begin by reviewing the definition of the quasi-stationary
distribution. Consider a continuous-time Markov process $X_t$ taking
values $n = 0, 1, 2,...,S$, with the state $n\!=\!0$ absorbing. (In
any realization of the process, if $X_t = 0$, then $X_{t'}$ must be
zero for all subsequent times, $t' > t$.) The transition rates
$w_{m,n}$ (from state $n$ to state $m$) are such that $w_{m,0} = 0$,
$\forall m > 0$. Let $p_n(t)$ denote the probability of state $n$ at
time $t$, given some initial state $X_0$.   The {\it survival
probability} $P_s(t) = \sum_{n \geq 1} p_n(t)$ is the probability
that the process has not become trapped in the absorbing state up to
time $t$. We suppose that as $t \to \infty$ the $p_n$, normalized by
the survival probability $P_s(t)$, attain a time-independent form.
The {\it quasi-stationary distribution} $\overline{p}_n$ is then
defined via

\begin{equation}
\overline{p}_n  = \lim_{t \to \infty} \frac{p_n (t) }{P_s (t)}
,\;\;\;\; (n \geq 1),
\label{qshyp}
\end{equation}
with $\overline{p}_0 \equiv 0$. The QS distribution is
normalized so:
\begin{equation}
\sum_{n \geq 1} \overline{p}_n = 1.
\label{norm}
\end{equation}

The basis for our simulation method is the observation
\cite{RDVIDIGAL} that the QS distribution represents a stationary
solution to the equation,

\begin{equation}
\frac{d q_n}{dt} = -w_n q_n + r_n + r_0 q_n \;\; (n>0)\;,
\label{qme}
\end{equation}
where $w_n = \sum_m w_{m,n}$ is the total rate of transitions out of state
$n$, and $r_n = \sum_m w_{n,m} q_m$ is the flux of probability into this
state.  To see this, consider the master equation (Eq. (\ref{qme})
without the final term) in the QS regime.  Substituting
$q_n(t) = P_s(t) \overline{p}_n$, and noting that in the QS regime
$d P_s/dt = -\overline{r}_0 = - P_s \sum_m w_{0,m} \overline{p}_m$,
we see that the r.h.s. of Eq. (\ref{qme}) is identically zero
if $q_n = \overline{p}_n$ for $n \geq 1$.
The final term in Eq. (\ref{qme}) represents a redistribution of the
probability $r_0$ (transfered to the
absorbing state in the original master equation),
to the nonabsorbing subspace.
Each nonabsorbing state receives a share
equal to its probability.

Although Eq. (\ref{qme}) is not a master equation (it is nonlinear
in the probabilities $q_n$), it does suggest a simulation scheme for
sampling the QS distribution. In a Monte Carlo simulation one
generates a set of realizations of a stochastic process.  In what
follows we call a simulation of the original process $X_t$
(possessing an absorbing state) a {\it conventional} simulation. We
define a related process, $X_t^*$, whose {\it stationary}
probability distribution is the {\it quasi-stationary} distribution
of $X_t$. (Note that in order to have a nontrivial stationary
distribution, $X_t^*$ cannot possess an absorbing state.) The
probability distribution of $X_t^*$ is governed by Eq. (\ref{qme}),
which implies that for $n>0$ (i.e., away from the absorbing state),
the evolution of $X_t^*$ is identical to that of $X_t$. (Since  Eq.
(\ref{qme}) holds for $n > 0$, the process $X_t^*$ must begin in a
non-absorbing state.) When $X_t$ enters the absorbing state,
however, $X_t^*$ instead jumps to a nonabsorbing one, and then
resumes its ``usual" evolution (i.e., with the same transition
probabilities as $X_t$), until such time as another visit to the
absorbing state is imminent.  The probability that $X_t^*$
jumps to state $j$ (when a visit to state 0 is imminent),
is the QS probability $\overline{p}_j$.

A subtlety associated with this procedure is that the QS distribution
is needed to determine the evolution of $X_t^*$ when $X_t$ visits
the absorbing state.  Although one has no prior knowledge of
the QS distribution $\overline{p}_n$, one
can, in a simulation, use the history $X_s^*$ ($0 < s \leq t$) up to
time $t$, to {\it estimate} the $\overline{p}_n$.  (There is good evidence,
after all, that the surviving sample in a conventional simulation
converges to the QS state after an initial transient.) In practice
this is accomplished by saving (and periodically updating) a sample
$n_1, n_2, ..., n_M$ of the states visited. As the evolution
progresses, $X_s^*$ will visit states according to the QS
distribution. We therefore update the sample $\{n_1,n_2,...,n_M\}$
by occasionally replacing one of these configurations with the
current one. In this way the distribution for the process $X_t^*$
(and the sample drawn from it), will converge to the QS distribution
(i.e., the stationary solution of Eq. (\ref{qme})) at long times.
Summarizing, the simulation process $X_t^*$ has the same dynamics as
$X_t$, except that when a transition to the absorbing state is
imminent, $X_t^*$ is placed in a nonabsorbing state, selected at
random from a sample over the history of the realization. In effect,
the nonlinear term in Eq. (\ref{qme}) is represented as a {\it
memory} in the simulation.

\subsection{The contact process}

To explain how our method works in practice, we detail its
application to the {\it contact process} (CP)
\cite{harris,liggett,marro}. In the CP, each site $i$ of a lattice
is either occupied ($\sigma_i (t)= 1$), or vacant ($\sigma_i (t)=
0$).  Transitions from $\sigma_i = 1$ to $\sigma_i = 0$ occur at a
rate of unity, independent of the neighboring sites. The reverse
transition is only possible if at least one neighbor is occupied:
the transition from $\sigma_i = 0$ to $\sigma_i = 1$ occurs at rate
$\lambda r$, where $r$ is the fraction of nearest neighbors of site
$i$ that are occupied; thus the state $\sigma_i = 0$ for all $i$ is
absorbing. ($\lambda $ is a control parameter governing the rate of
spread of activity.)

Although no exact results are available, the CP has been studied
intensively via series expansion and Monte Carlo simulation.  The
model has attracted much interest as a prototype of a
nonequilibrium critical point, a simple representative of the
directed percolation (DP) universality class.  Since its scaling
properties have been discussed extensively
\cite{marro,hinrichsen,odor04} we review them only briefly.  The
best estimate for the critical point in one dimension is $\lambda_c
= 3.297848(20)$, as determined via series analysis \cite{iwanrd93}.
As the critical point is approached, the correlation length $\xi$
and correlation time $\tau$ diverge, following $\xi \propto
|\Delta|^{-\nu_\perp}$ and $\tau \propto |\Delta|^{-\nu_{||}}$,
where $\Delta = (\lambda - \lambda_c)/\lambda_c$ is the relative distance from the
critical point.  The order parameter (the fraction of active sites),
scales as $\rho \propto \Delta^\beta$ for $\Delta > 0$.

Two characteristic times, $\tau_C$ and $\tau_L$, may be identified
in the contact process.  The first is a relaxation time that governs
the decay of temporal correlations in the stationary state: $C(t)
\equiv \langle \rho(t_0) \rho(t_0+t) \rangle - \rho^2 \sim
e^{-t/\tau_c}$. The second is a lifetime, determining the asymptotic
decay of the survival probability (starting from a spatially
homogeneous initial condition) via $P(t) \sim e^{-t/\tau_L}$. The
two characteristic times exhibit the same scaling properties in the
critical region. In the supercritical or active phase ($\Delta >
0$), the lifetime grows exponentially with system size and $\Delta$,
while $\tau_C$ remains finite. Our interest here is in the lifetime
(denoted simply as $\tau$) of the quasi-stationary state.  In the QS
probability distribution there is a nonzero flux of probability to
the absorbing state,

\begin{equation}
r_0 = w_{01} \overline{p_1} \label{r0}
\end{equation}

\noindent where $\overline{p_1}$ is the QS probability of the
configuration with exactly one active site and $w_{01} = 1$ is the
transition rate from this configuration to the absorbing state; the
QS lifetime $\tau = 1/r_0$.  (In QS simulations we take $\tau$ to be
the mean time between successive attempts to visit to the absorbing
state.)

An important point in interpreting our simulation results concerns
the finite-size scaling (FSS) behavior of $\tau$.  According to the
usual FSS hypothesis \cite{fisherfss}, finite-size corrections to
critical properties are functions of the ratio $L/\xi$, or,
equivalently, of the quantity $\Delta L^{1/\nu_\perp}$.  The
lifetime is therefore expected to follow (in the subcritical phase,
$\Delta < 0$),

\begin{equation}
\tau (\Delta,L) = |\Delta|^{-\nu_{||}} {\cal F}(|\Delta|
L^{1/\nu_\perp}) \label{taufss}
\end{equation}

\noindent where the scaling function ${\cal F}(x) \propto
x^{\nu_{||}}$ for small $x$ (so that $\tau$ does not diverge in a
finite system), while in the opposite limit ${\cal F} \to {\cal
F}_0$, a constant. The scaling hypothesis leads to the familiar
relation, $\tau(0,L) \sim L^{\nu_{||}/\nu_\perp}$ at the
critical point, and suggests that we attempt to collapse data for
diverse system sizes by plotting $\Delta^{\nu_{||}} \tau$ versus
$\Delta^* \equiv \Delta L^{1/\nu_\perp}$.  For the order parameter the expected finite-size
scaling form is \cite{marro},

\begin{equation}
\rho(\Delta,L) = |\Delta|^\beta {\cal R} (L^{1/\nu_\perp} \Delta) \;.
\label{fssrho}
\end{equation}
In the subcritical regime, the order parameter must decay to zero
$\propto L^{-1}$ as $L \to \infty$, for any $\Delta < 0$, so that
${\cal R} (x) \sim |x|^{-\nu_\perp}$ as $x \to -\infty$.  On the
other hand, for $\Delta = 0$ and $L$ finite, $\rho$ must be nonzero
and nonsingular, implying ${\cal R}(x) \sim x^{-\beta}$ for $x \to
0$.   Thus $\rho \sim |\Delta|^{\beta - \nu_\perp}$ for $\Delta$
large and negative.

\section{Simulation results}

In the QS simulations we use a list of size $M= $ 2 $\times 10^3$ -
$10^4$, depending on the lattice size. The process is simulated in
15 realizations, each of $5 \times 10^8$ time steps. As is usual,
annihilation events are chosen with probability $1/(1+\lambda)$ and
creation with probability $\lambda/(1+\lambda)$.  A site $i$ is
chosen from a list of currently occupied sites, and, in the case of
annihilation, is vacated.  In a creation event, a nearest-neighbor
of site $i$ is selected at random and, if it is currently vacant, it
becomes occupied.  The time increment associated with each event is
$\Delta t = 1/N_{occ}$, where $N_{occ}$ is the number of occupied
sites just prior to the attempted transition \cite{marro}.

In the initial phase of the evolution, the list of saved
configurations is augmented whenever the time $t$ increases by one,
until a list of $M$ configurations has been accumulated. From then
on, we update the list (replacing a randomly selected entry with the
current configuration), with a certain probability $p_{rep}$,
whenever $t$ advances by one unit. A given configuration therefore
remains on the list for a mean time of $M/p_{rep}$.  (Values of
$p_{rep}$ in the range $10^{-3} - 10^{-4}$ are used.)

Fig. 1 shows the QS lifetime $\tau$ as a function of $\Delta$,
for lattice sizes $L= 20$, 40, 80,...,2560.  For the larger system
sizes, power-law dependence on $\Delta$ is evident, before the
lifetime saturates (at very small values of $|\Delta|$), due, as
anticipated, to finite-size effects. In Fig. 2 these data are
collapsed using the known values of the DP exponents \cite{jensen99}, $\nu_\perp =
1.09684(6)$ and $\nu_{||} = 1.73383(3)$.  The data collapse is
quite good for the larger system sizes.  A least-squares fit to the
data in the linear portion of the graph ($|\Delta| \geq 0.02$)
yields a slope of -1.738(12), in reasonable agreement with the accepted
value for $\nu_{||}$.
An alternative method for analyzing the data is to estimate, for each value
of $\Delta$, the infinite-size limiting value of the lifetime by plotting
$\tau(\Delta,L)$ versus $1/L$ (see Fig. 3).  Plotting the resulting estimates
on log scales, we find a slope of -1.735(9), again in good agreement with
the standard value for the exponent $\nu_{||}$.

We verified that the distribution $p_T$ of the lifetime (that is, of the time interval
between successive attempts to visit the absorbing state) is {\it exponential}:

\begin{equation}
p_T(t) = \frac{e^{-t/\tau}}{\tau}
\end{equation}
This is as expected: in the quasi-stationary state the transition
rate to the absorbing state is time-independent.

Turning to the order parameter, we see from Eq. (\ref{fssrho}) that
a plot of $\rho^* \equiv |\Delta|^{-\beta} \rho$ versus $\Delta^* $
should exhibit a data collapse.  The asymptotic behavior of the
scaling function ${\cal R}$ implies that $\rho^* \propto
(\Delta^*)^{-\nu_\perp}$ for large $\Delta^*$. These scaling
properties are verified in Fig. 4. (We use the accepted value $\beta
= 0.2765$. While scaling of the order parameter in the subcritical
regime was verified in Ref \cite{marro}, here we are able to extend
the range of $\Delta^*$ by an order of magnitude, using the QS
simulation technique.)

In Fig. 5 we show the probability distribution $p(\rho)$ of the order parameter in a large
system ($L=2560$) for three values of $\Delta$.  As expected, the distribution broadens and
shifts toward larger values of $\rho$ as $\lambda$ approaches the critical value.
The probability distribution follows, to a good approximation, the scaling form

\begin{equation}
p(\rho) = \frac{1}{\langle \rho \rangle} {\cal P} (\rho/\langle \rho \rangle)
\end{equation}
where $\langle \rho \rangle \equiv \rho(\Delta,L) $ is the mean value and
${\cal P}$ is a scaling function.  This is verified (Fig. 6) by plotting
$p^* =  \langle \rho \rangle p(\rho)$ versus $\rho/\langle \rho \rangle$.
We see that the scaling function ${\cal P}$ attains its maximum near
$\rho^* = 0.6$, and that it falls off rapidly as $\rho^* \to 0$.  On the other side
of the maximum it exhibits a roughly exponential tail.

\section{Summary}

We use the quasi-stationary simulation method to study the
lifetime and order parameter of the one-dimensional contact process in
the subcritical phase.
Our results confirm the expected scaling properties of the lifetime of the
quasi-stationary state, and of the order parameter.
 The QS simulation method is the
first to allow such an analysis deep in the subcritical regime. With
a rather modest expenditure of computer time, the approach yields an
estimate of the critical exponent $\nu_{||}$ that agrees with the
accepted value to within uncertainty, with a precision of about
0.5\%.  Analysis of quasi-stationary properties in the subcritical
regime should therefore be a useful tool in the study of
absorbing-state phase transitions.

\vspace{1em}

\noindent{\bf Acknowledgment}

This work was supported by CNPq and FAPEMIG, Brazil.

\bibliographystyle{apsrev}

\newpage

FIGURE CAPTIONS
\vspace{1em}

\noindent FIG. 1. QS survival time $\tau$ as a function of $\Delta$,
for lattice sizes $L= 20$, 40, 80,...,2560.
The slope of the straight line is -1.74.
\vspace{1em}

\noindent FIG. 2. Data of Fig.1 plotted in terms of the scaling variables
$\Delta^* =  L^{1/\nu_\perp} |\Delta|$ and $\tau^* = L^{-z} \tau$, with $z =
\nu_{||}/\nu_\perp$.
The slope of the straight line is -1.734.
\vspace{1em}

\noindent FIG. 3. QS survival time $\tau$ as a function of $\Delta$,
for $|\Delta|= 0.3, 0.2, 0.1, 0.05, 0.02, 0.01, 0.005, 0.002, 0.001,
0.0005$, from bottom to top (lattice sizes $L= 20$, 40,
80,...,2560). Lifetime by plotting $\tau(\Delta,L)$ versus $1/L$.
Inset: infinite-size limiting estimates of the lifetime as function
of $|\Delta|$, plotted on log scales.

\noindent FIG. 4. Scaled QS order parameter
$\rho^* \equiv |\Delta|^{-\beta} \rho$ versus
$\Delta^* = L^{1/\nu_\perp} |\Delta|$
for $L=320$ (open squares), $L=1280$ (diamonds) and $L=2560$
(filled squares).  The slope of the
straight line is  -1.1097.

\noindent FIG. 5. QS probability distribution of the order parameter
in the subcritical regime, for $L=2560$ and (left to right)
$\lambda/\lambda_c = 0.8$, 0.9 and 0.95.

\noindent FIG. 6. Scaling plot of the data shown in Fig. 5;
$\lambda/\lambda_c = 0.8$ (squares), 0.9 (circles) and 0.95 (triangles).
\vspace{1em}

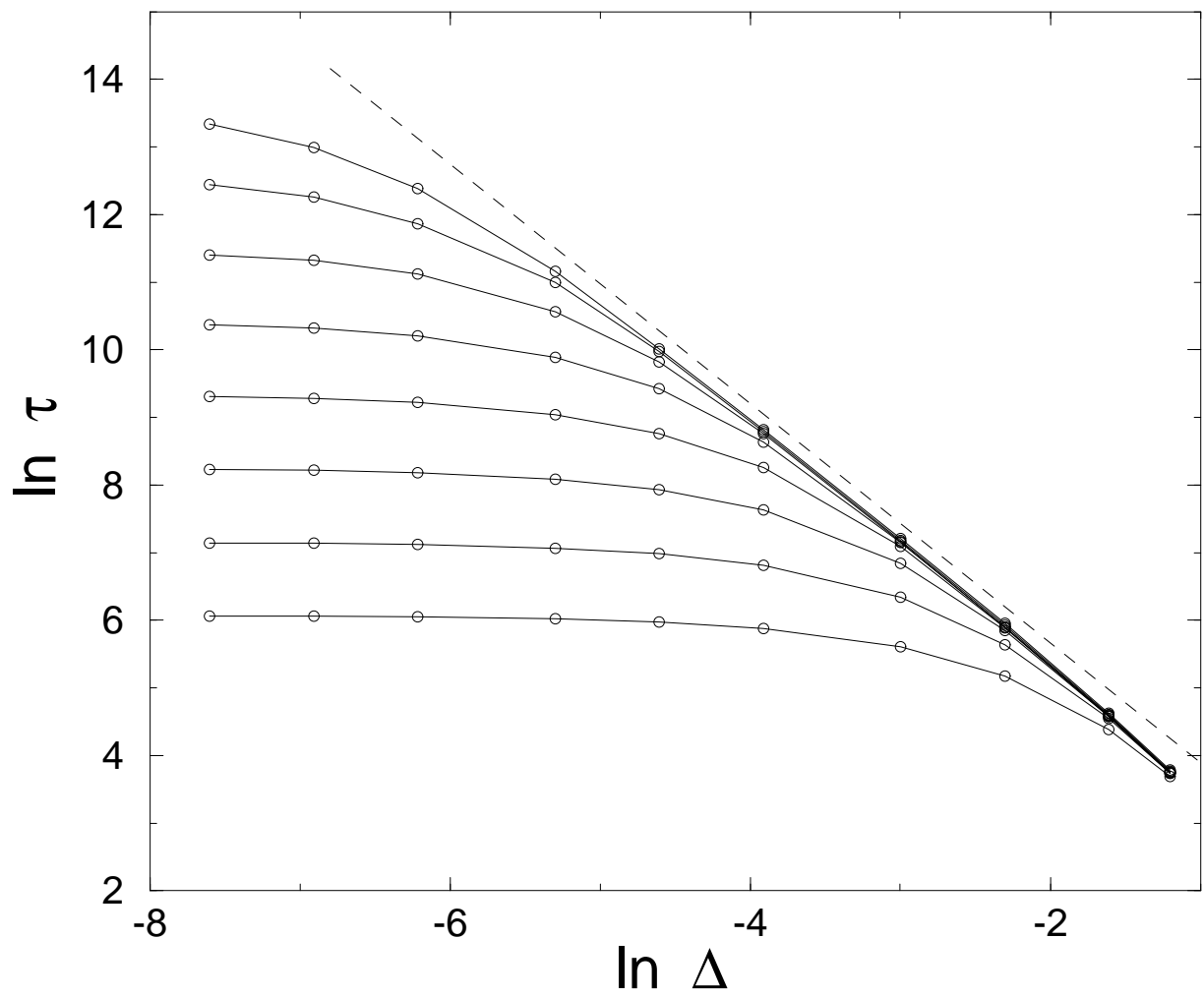
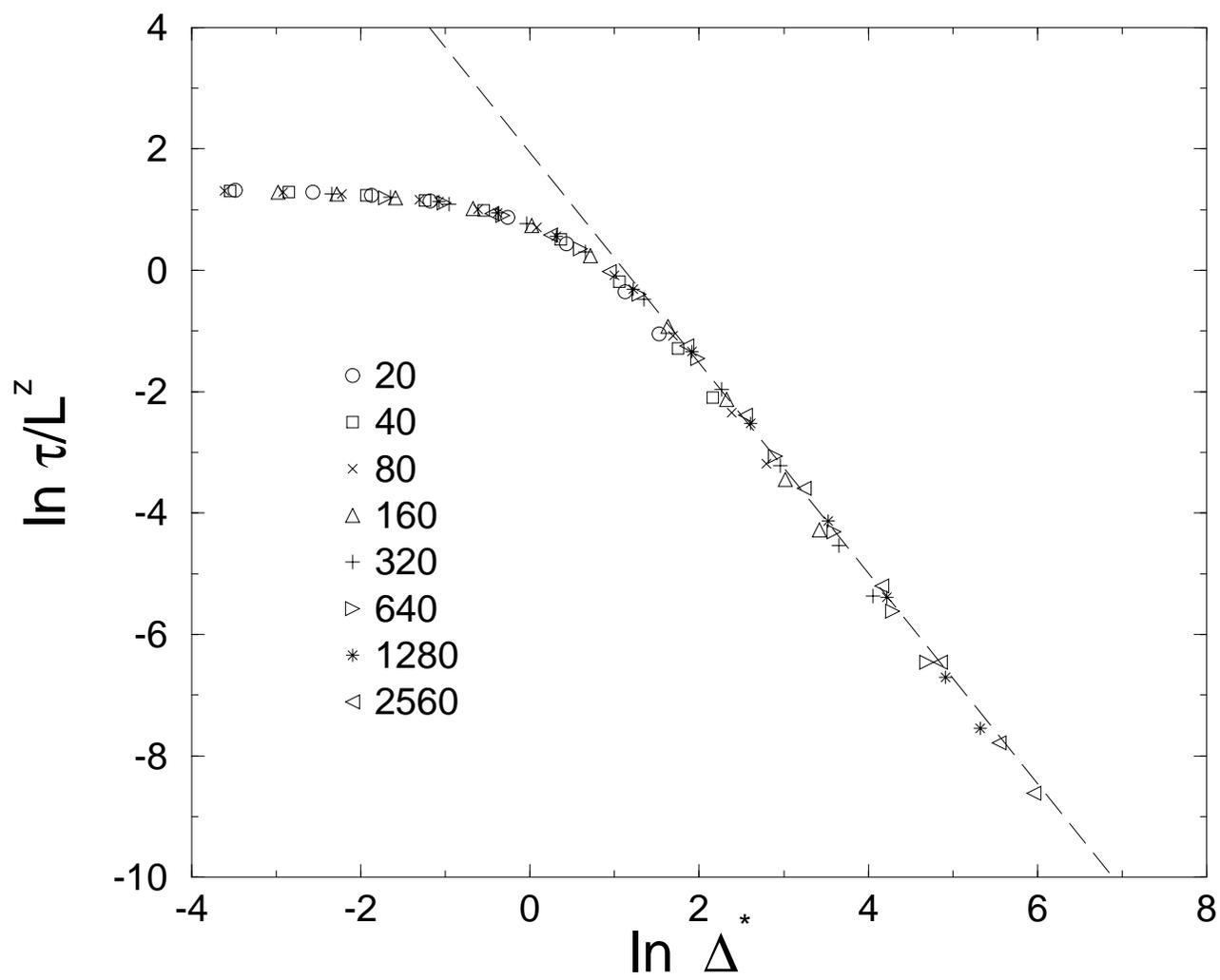
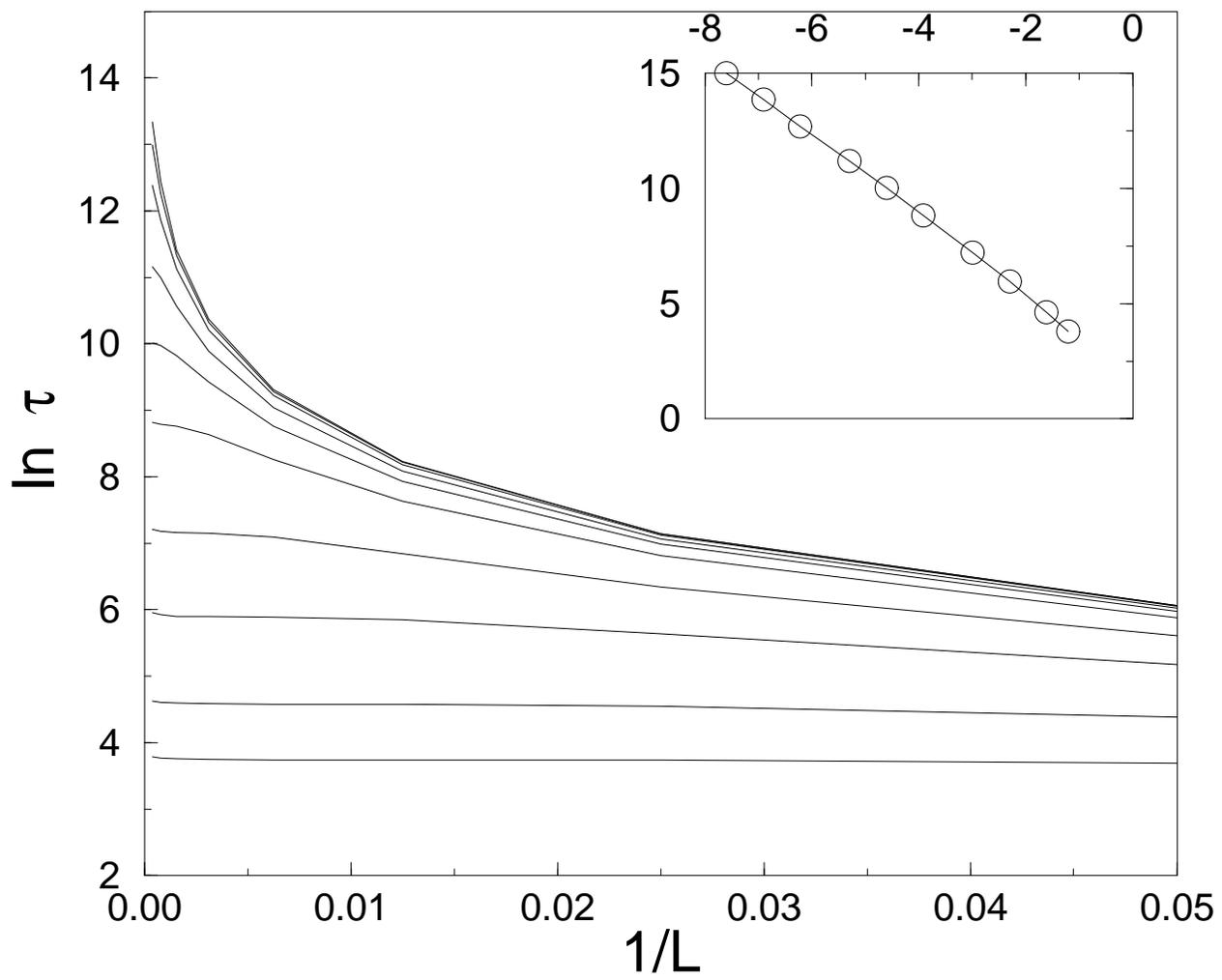
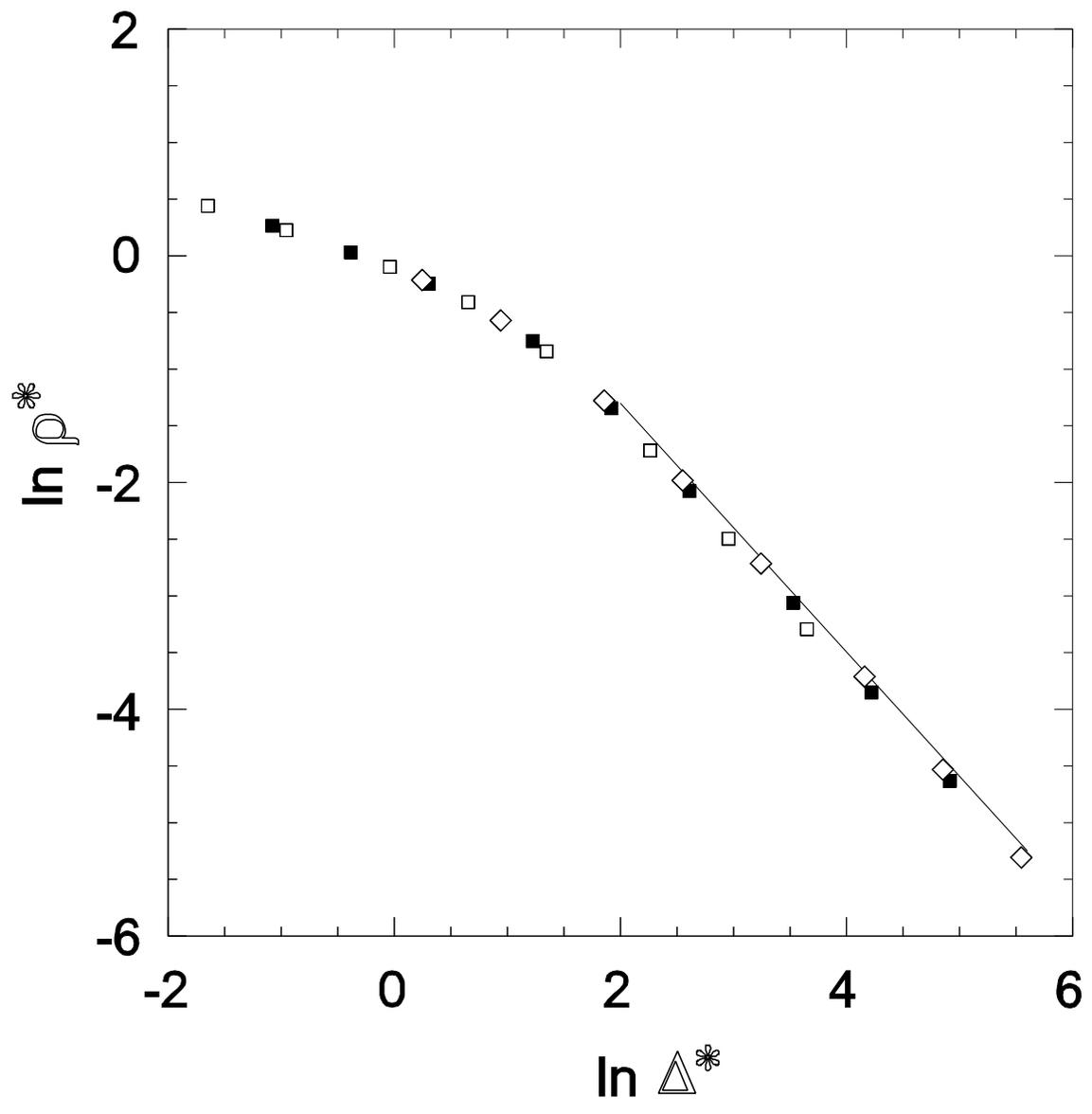
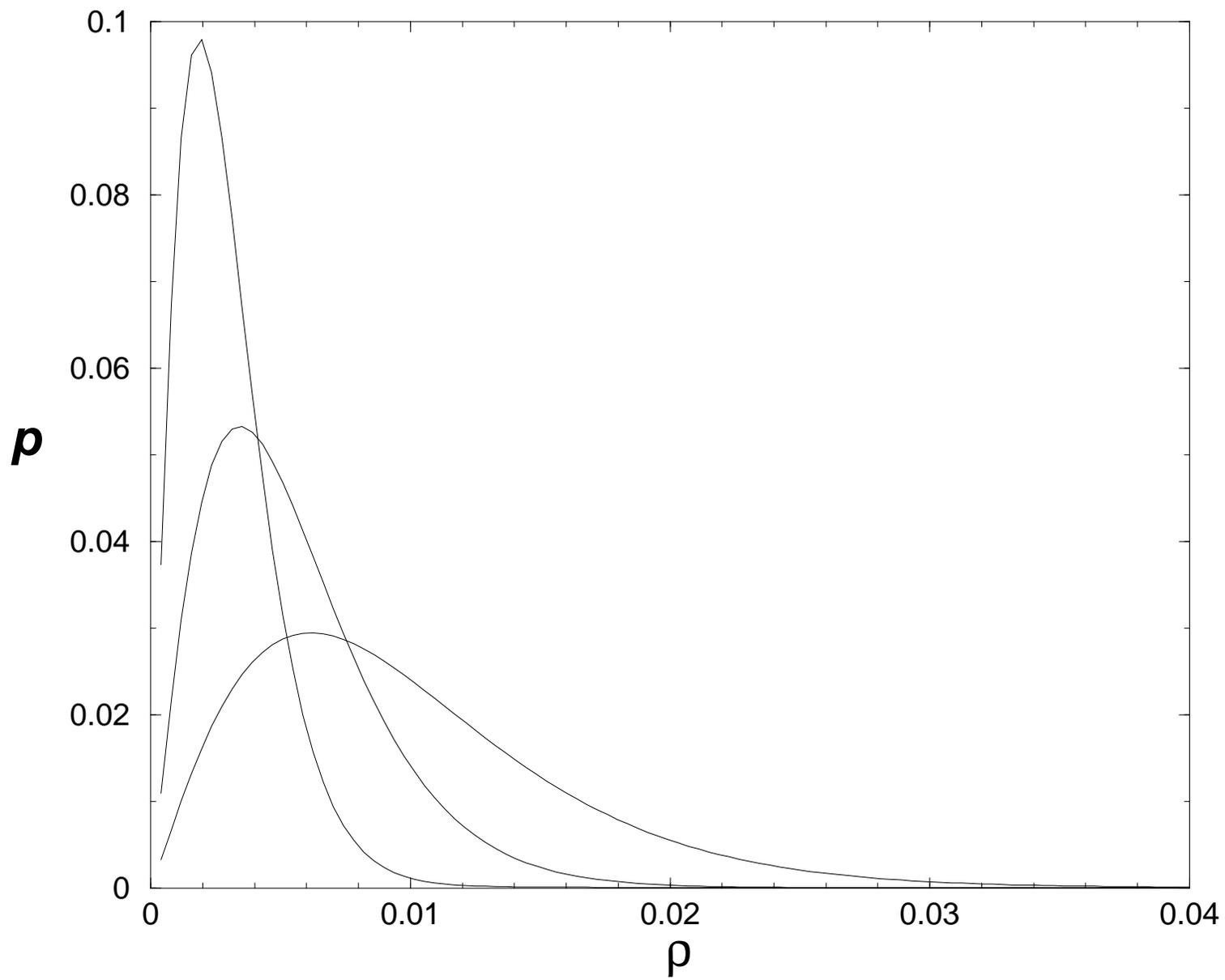
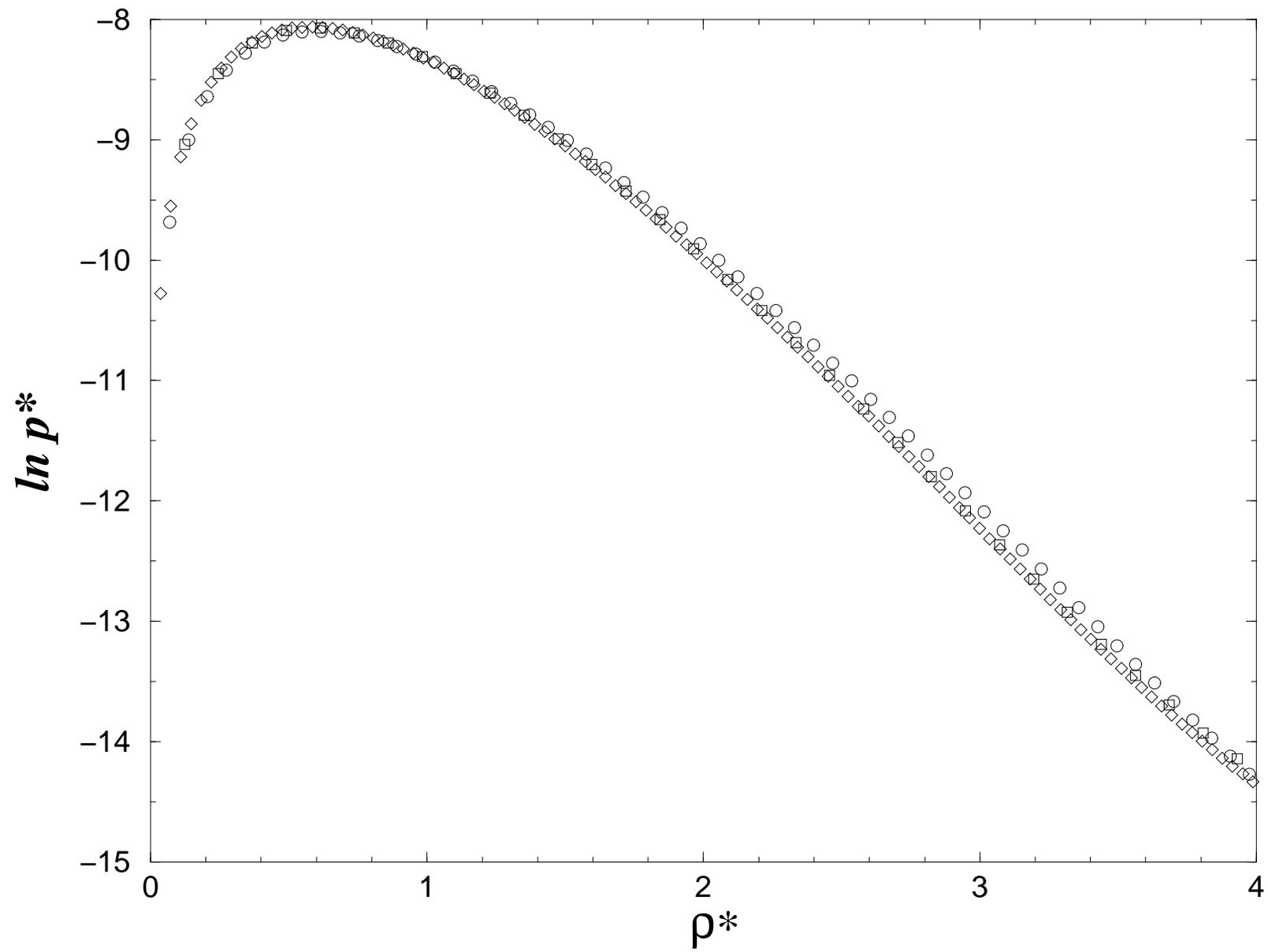
\end{document}